\documentclass[trackchanges]{aastex701}
\usepackage{comment}
\usepackage{savesym}
\savesymbol{tablenum}
\usepackage{amsmath}
\usepackage{siunitx}
\restoresymbol{SIX}{tablenum}
\usepackage{longtable}
\usepackage{hyperref}
\begin{document}

\title{Is the Dark Comet 1998 KY$_{26}$ the Spacecraft Phobos 1?}

\author[orcid=0000-0003-1116-576X,sname='Hibberd']{Adam Hibberd}
\altaffiliation{i4is}
\affiliation{Initiative for Interstellar Studies, 27/29 South Lambeth Road London, SW8 1SZ United Kingdom}
\email[show]{adam.hibberd@i4is.org}  

\author{Adam Crowl}
\altaffiliation{i4is}
\affiliation{Initiative for Interstellar Studies, 27/29 South Lambeth Road London, SW8 1SZ United Kingdom}
\email[show]{adam.crowl@i4is.org}

\author[orcid=0009-0007-3343-2001,sname='Gómez de Olea B.']{Carlos Gómez de Olea Ballester}
\altaffiliation{TUM}
\affiliation{Initiative for Interstellar Studies, 27/29 South Lambeth Road London, SW8 1SZ United Kingdom}
\affiliation{Technical University of Munich (TUM), Arcisstraße 21, 80333 Munich, Germany}
\email[show]{carlos.olea@tum.de} 
\author[orcid=0000-0003-4330-287X,sname='Loeb']{Abraham Loeb}
\affiliation{Astronomy Department, Harvard University, 60 Garden Street, Cambridge MA 02138 USA}
\email[show]{aloeb@cfa.harvard} 

\begin{abstract}

Since the discovery of new kinds of celestial bodies known as \textit{dark comets}, scientists have speculated about their ontology. A curious hybrid of comet and asteroid, these objects show significant non-gravitational accelerations (NGAs) yet exhibit absolutely no signs of cometary outgassing in the form of a coma or tail. The planned rendezvous of the Hayabusa2 spacecraft with 1998 KY$_{26}$ in July 2031 elevates the question of this so-called dark comet’s nature beyond a purely research exercise, as the true nature of the object may have practical implications for the scientific return of the mission. This study examines the hypothesis that 1998 KY$_{26}$ may be of technogenic origin, in fact a relic of a historical Russian mission to Mars, the Phobos 1 probe, which suffered a failure 2 months after the launch in July 1988, due to upload of a faulty command. We find that two propulsive $\Delta$Vs combined at 1.9 \si{km.s^{-1}}, the first just after loss of mission and the second in May 1996, allow the orbits and phases of the two bodies to align, with an arbitrarily low \textit{Mahalanobis distance} using the covariance of the dark comet in 6D phase space. There is also evidence that 1.9 \si{km.s^{-1}} was within the performance envelope of Phobos 1, which had a powerful nitric acid and amine-based autonomous thruster for Mars Orbital Insertion (MOI).


\end{abstract}



\section{Introduction}

Ever since \cite{2023LPICo2851.2036S} identified the existence of celestial bodies with orbits perturbed by significant non-gravitational accelerations (NGAs) - (A$_1$, A$_2$, A$_3$), radial, transverse and  perpendicular to the orbital plane respectively \citep{1973AJ.....78..211M}, yet exhibiting no signs of outgassing; the existence and nature of these objects have been the subject of hot debate. Several further such objects were quickly discovered and presented in \cite{Seligman_2024} which helpfully delineated them into two distinct populations, the \textit{Outer Population} (those with Jupiter Family Comet (JFC) orbits,  and the \textit{Inner Population} (with lower semi-major axes and eccentricities).\\

So what IS the source of these anomalous NGAs? In an attempt to narrow the limits on dust ejection, the Japanese Aerospace Exploration Agency (JAXA) will be conducting an analysis into one dark comet, 1998 KY$_{26}$ by sending their \textit{Hayabusa2} spacecraft to rendezvous \citep{HIRABAYASHI20211533} with it in 2031. The instrumentation should hopefully be able to detect low levels of dust particle generation possibly sufficient to ascertain whether the rocket effect can indeed be the cause of the non-gravitational acceleration, but more importantly we will finally know what the object actually looks like.\\

Dark comet 1998 KY$_{26}$ has been studied extensively in the scientific literature \citep{ostro1999radar,santana2025hayabusa2,bolin2025keck,beniyama2025size,ostro19981998,pravec19981998,tholen2003recovery,hicks1998close,farnocchia2025radiation}. In summary, it is a small ($\sim{11}$ $\si{m}$), extremely rapidly rotating (rotation period $\sim{5.3}$ min) object with an unusally high albedo ($\sim{0.52}$).\\

Ongoing investigations and previous research by authors of this paper (e.g. \cite{loeb2025dark}), suggest that dark comets could actually be derelict spacecraft missions, and the NGAs could be the result of solar radiation pressure.\\

The report herein addresses the possible nature of 1998 KY$_{26}$ and proposes a technogenic (man-made) origin for this object, specifically the Russian Phobos 1 failed mission to Mars, launched 07/07/1988 \citep{Phobos1_2}. For a description and history of these probes, the most massive interplanetary spacecraft ever launched, go to \cite{sagdeev1990brief}.\\

\section{Supporting Evidence}

A comprehensive computation of all interplanetary missions since the start of the space age was conducted by repeatedly  executing Optimum Interplanetary Trajectory Software - OITS \citep{OITS_info,AH2,HPH21} - with different mission-specific input parameters. Developed by Adam Hibberd, OITS has been exploited for research into missions to interstellar objects, but also to celestial bodies belonging to our own Solar System \citep{HPE19,HEL22,HHE20,HH21,AH23,HA23}. Thus an accurate determination of the trajectory profile for each historical mission could be established. The trajectory of 1998 KY$_{26}$ could then be compared in various ways with each of these missions in turn to establish any similarities. 

\subsection{Comparison of Orbital Elements}

The osculating orbital elements of 1998 KY$_{26}$ as of 01/01/2001 are shown in Table \ref{tab:DCorb} - refer to row 8.\\

Observe that the perihelion, $q = 0.984$ $\si{au}$ is close to Earth's orbit and the aphelion $Q = 1.482$ $\si{au}$, is around the orbit of Mars. Furthermore the orbital plane is close to the ecliptic with an inclination of 1.48 $^{\circ}$, slightly less than that of the orbit of Mars ($\sim{1.9}$ $^{\circ}$). This has similarities to a trajectory a mission planner might design to reach Mars, thus clearly it would be most pertinent to compare its orbit with previous Mars missions.\\

Since there are 5 orbital parameters needed to specify an orbit's shape, size and orientation, we must find a way of comparing one orbit with another in order to determine the 'net orbital difference'.\\

To this end, we adopt the so-called non-singular orbital parameters defined as follows:

\begin{flalign}
    e1 = a,~~e2 = e cos\left(\Omega+\omega\right),~~ e3 = e cos\left(\Omega+\omega\right),~~ e4 = e sin\left(\Omega+\omega\right),~~ e5 = sin\left(\frac{i}{2}\right)cos(\Omega),~~    e6 = sin\left(\frac{i}{2}\right)sin(\Omega)
\end{flalign}

where the usual symbols are employed for the Keplerian orbital parameters. Thus for each comparison between 1998 KY$_{26}$ and a mission, the net orbital difference can be taken as the norm (or 'root sum square') of the difference in the 6 non-singular components, $e1,~ e2,~ e3,~ e4,~ e5,~ e6$.\\

\begin{figure}[hbt!]
\centering
\includegraphics[width=0.9\textwidth]{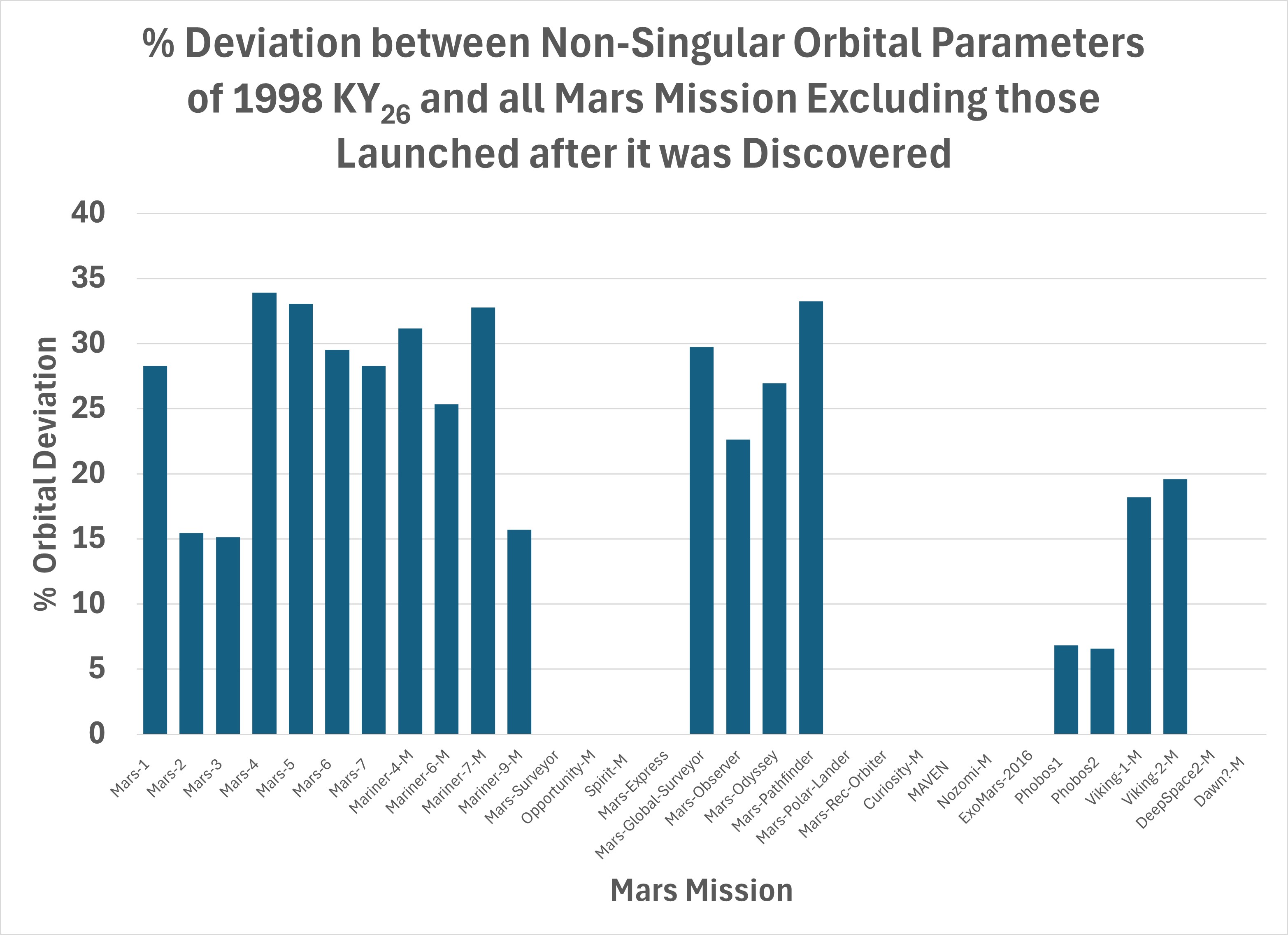}
\caption{Magnitude vector difference in non-singular orbital parameters between 1998 KY$_{26}$ and 31 historical Mars missions.}
\label{fig:Orbd}
\end{figure}

 Figure \ref{fig:Orbd} represents the results of this analysis.\\
\begin{table}
\centering
\begin{tabular}{|c|c|c|c|c|c|c|c|c|c|}
\hline
\textbf{Number} &
  \textbf{Designation} &
  \textbf{Discovery} &
  \textbf{H (mag)} &
  \textbf{a (\si{au})} &
  \textbf{q (\si{au})} &
  \textbf{e} &
  \textbf{i (deg)} &
  \textbf{$\Omega$ (deg)} &
  \textbf{$\omega$ (deg)} \\ \hline
\textbf{1}  & \textbf{2001 ME$_1$}   & 25/05/2001 & 16.53 & 2.648 & 0.353 & 0.867 & 5.811  & 86.212  & 300.523 \\ \hline
\textbf{2}  & \textbf{2005 UY$_6$}   & 29/10/2005 & 18.14 & 2.257 & 0.294 & 0.870 & 12.153 & 343.600 & 180.776 \\ \hline
\textbf{3}  & \textbf{1998 FR$_{11}$}  & 24/03/1998 & 16.42 & 2.812 & 0.827 & 0.706 & 6.660  & 129.909 & 158.522 \\ \hline
\textbf{4}  & \textbf{2012 UR$_{158}$} & 22/10/2012 & 20.7  & 2.239 & 0.324 & 0.855 & 3.220  & 287.738 & 238.091 \\ \hline
\textbf{5}  & \textbf{2003 RM}    & 02/09/2003 & 19.8  & 2.921 & 1.165 & 0.601 & 10.854 & 336.702 & 324.518 \\ \hline \hline
\textbf{6}  & 2005 VL$_1$   & 04/11/2005 & 26.45 & 0.892 & 0.691 & 0.225 & 0.247  & 39.812  & 226.461 \\ \hline
\textbf{7}  & 2010 RF$_{12}$  & 01/09/2010 & 28.42 & 1.061 & 0.861 & 0.188 & 0.883  & 163.838 & 267.583 \\ \hline
\textbf{8}  & 1998 KY$_{26}$  & 28/05/1998 & 25.6  & 1.233 & 0.984 & 0.202 & 1.481  & 84.365  & 209.372 \\ \hline
\textbf{9}  & 2016 NJ$_{33}$  & 28/05/2016 & 25.49 & 1.312 & 1.038 & 0.208 & 6.640  & 279.682 & 24.070  \\ \hline
\textbf{10} & 2010 VL$_{65}$  & 04/11/2010 & 29.22 & 1.069 & 0.915 & 0.144 & 4.713  & 223.635 & 252.126 \\ \hline
\textbf{11} & 2013 BA$_{74}$  & 31/01/2013 & 25.4  & 1.759 & 0.976 & 0.445 & 5.302  & 310.753 & 202.871 \\ \hline
\textbf{12} & 2006 RH$_{120}$ & 14/09/2006 & 29.5  & 0.996 & 0.960 & 0.036 & 0.283  & 49.644  & 121.622 \\ \hline
\textbf{13} & 2016 GW$_{221}$ & 14/04/2016 & 24.76 & 0.827 & 0.605 & 0.268 & 3.652  & 33.877  & 8.224   \\ \hline
\textbf{14} & 2013 XY$_{20}$  & 26/11/2013 & 25.65 & 1.131 & 1.010 & 0.106 & 2.863  & 78.634  & 18.187  \\ \hline
\end{tabular}
\caption{The dark comets as listed in \cite{Seligman_2024}, with magnitudes (extracted from therein) and osculating orbital elements as of 01/01/2001 taken from NASA JPL Horizons. The dark comets above the double horizontal bar and in bold are outer population and below are inner population}
\label{tab:DCorb}
\end{table}

We find the smallest discrepancy to be the Russian Mars mission 'Phobos 2', with the 'Phobos 1' mission following on close behind. Since Phobos 2 made it successfully to a Mars rendezvous, that would rule out this probe, whereas Phobos 1 was lost on an interplanetary trajectory after a faulty command was uploaded to it on 02/09/1988. \\
\begin{figure}[hbt!]
\centering
\includegraphics[width=0.9\textwidth]{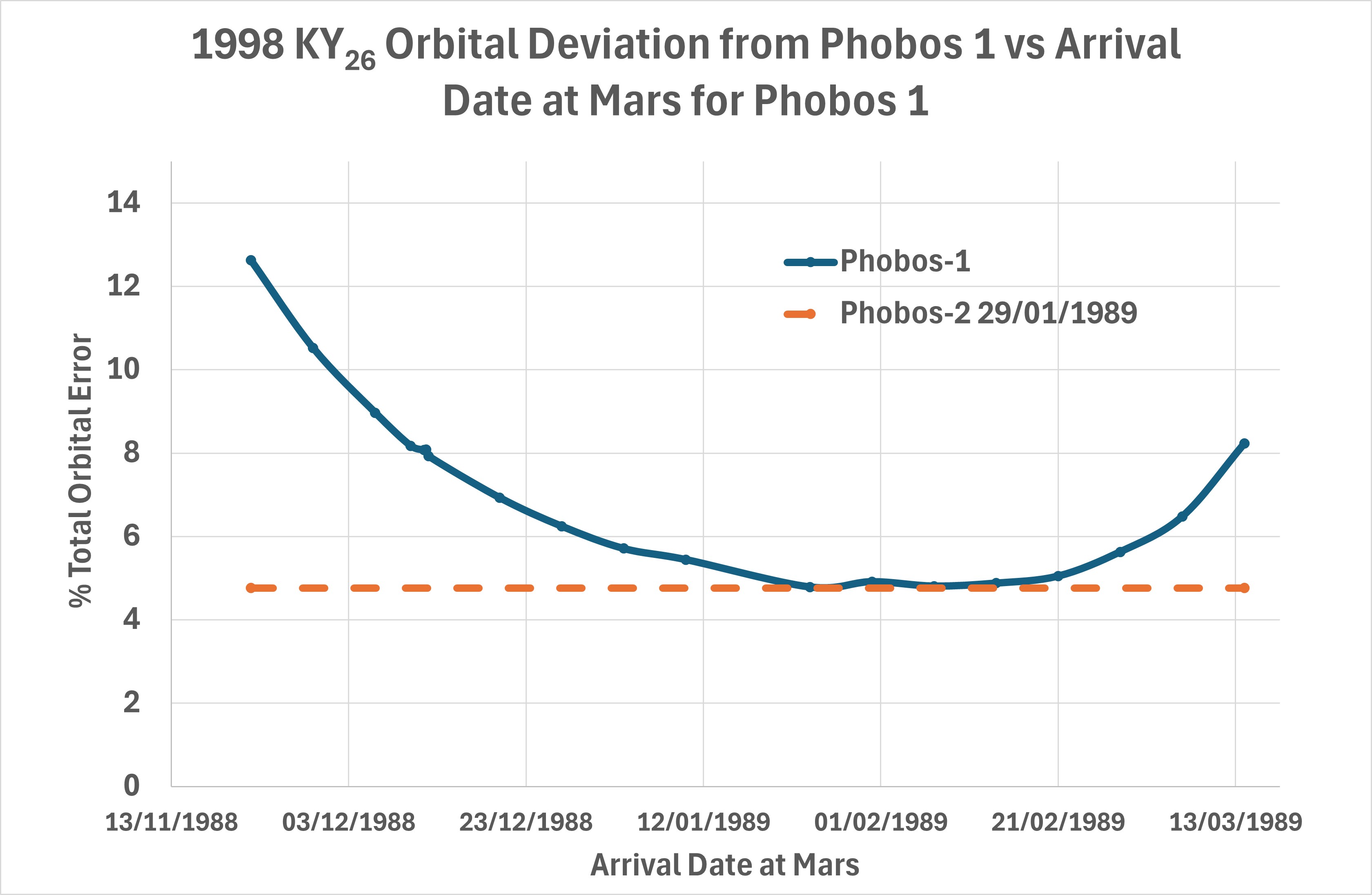}
\caption{Comparison of the orbital discrepancy between 1998 KY$_{26}$ and Phobos 1 as a dependency on possible arrival date at Mars, showing a minimum over a long period of arrival dates with the date adopted here 24/01/1988 a plausible assumption.}
\label{fig:OrbPhobos}
\end{figure}

Note that in order to properly ascertain the orbit of Phobos 1 (launched 07/07/1988), we need a reasonable notion of its planned arrival date, and there is some uncertainty over this. An arrival date of 24/01/1989 is adopted here, a few days before its twin and immediate successor Phobos 2 is known to have reached Mars. Refer Figure \ref{fig:OrbPhobos} which goes into further analysis of this.

\begin{table}[]
\centering
\begin{tabular}{|c|c|c|c|c|c|}
\hline
\textbf{}       & \textbf{}     & \textbf{Arrival}  & \textbf{Departure} & \textbf{$\Delta$V at}     & \textbf{Cumulative} \\
\textbf{Planet} & \textbf{Date} & \textbf{Velocity} & \textbf{Velocity}  & \textbf{Encounter} & \textbf{$\Delta$V}         \\
\textbf{}       & \textbf{}     & \textbf{\si{km.s^{-1}}}    & \textbf{\si{km.s^{-1}}}     & \textbf{\si{km.s^{-1}}}     & \textbf{\si{km.s^{-1}}}      \\ \hline
\textbf{Earth} & 1988 JUL 07 & 0.0000      & 3.4107 & 3.4107 & 3.4107 \\ \hline
\textbf{Mars}  & 1989 JAN 24 & 2.6349 & 0.0000      & 0.9881 & 4.3987 \\ \hline
\end{tabular}
\caption{Table of Velocities as Calculated by OITS for the 'Nominal' Phobos 1 Mission}
\label{Velocities}
\end{table}

\begin{table}[]
\centering
\begin{tabular}{|c|c|c|c|}
\hline
\textbf{Orbital   Parameter} & \textbf{Units} & \textbf{Phobos 1} & \textbf{1998 KY$_{26}$} \\ \hline
\textbf{Semi-major Axis, a} & au        & 1.302  & 1.233   \\ \hline
\textbf{Eccentricity, e}    & \textbf{} & 0.219  & 0.202   \\ \hline
\textbf{Perihelion, q}      & au        & 1.0166 & 0.984   \\ \hline
\textbf{Inclination, i}     &  $^{\circ}$         & 1.36   & 1.481   \\ \hline
\textbf{LOAN, $\Omega$}              &   $^{\circ}$     & -74.68 & 84.365  \\ \hline
\textbf{AOP, $\omega$}               &    $^{\circ}$   & 368.31 & 209.372 \\ \hline
\end{tabular}
\caption{Orbital elements of Phobos 1 in Comparison to 1998 KY$_{26}$}
\label{Orbits}
\end{table}

\subsection{Visual Comparison of Orbits}

\begin{figure}[hbt!]
\centering
\includegraphics[width=0.9\textwidth]{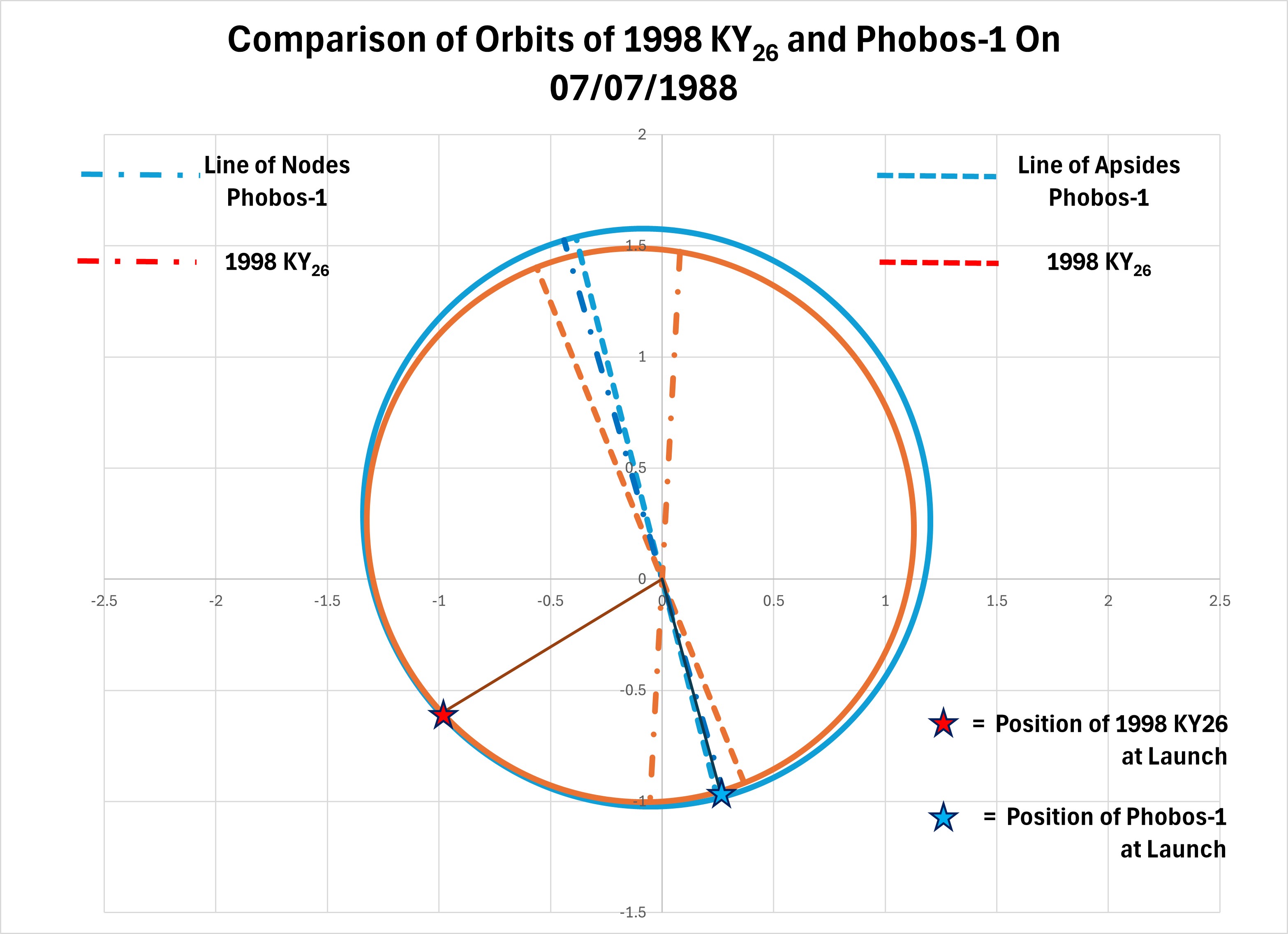}
\caption{2D view of the orbits of 1998 KY$_{26}$ and the Phobos 1 probe at the time of launch, showing a striking similarity between the two orbits.}
\label{fig:Orbdepict}
\end{figure}

The orbits of Phobos 1 and 1998 KY$_{26}$ as of the launch of the mission are depicted in Figure \ref{fig:Orbdepict}. Observe there is an obvious similarity between these two orbits, albeit the locations of the two objects are separated by a significant difference in ecliptic longitude, and there is a noticeable discrepancy in semi-major axis, a.

\subsection{Comparison of Diameter}\label{diam}

Assuming a span for each solar panel of $\sim{5}$ m, that would give the overall diameter of $\sim{10}$ m. From \cite{santana2025hayabusa2}, 1998 KY$_{26}$ has a diameter of $11 \pm 2$ m.
\subsection{Comparison of Brightness}

The absolute magnitude of 1998 KY$_{26}$ from Table \ref{tab:DCorb} is 25.6 and taken from \cite{Seligman_2024}. Assuming a highly reflective albedo for the Phobos probes of 1, with the diameter adopted in Section \ref{diam}, we get a calculated absolute magnitude of 25.5, which is in close agreement.

\subsection{Light Curve}
\begin{figure}[hbt!]
\centering
\includegraphics[width=0.9\textwidth]{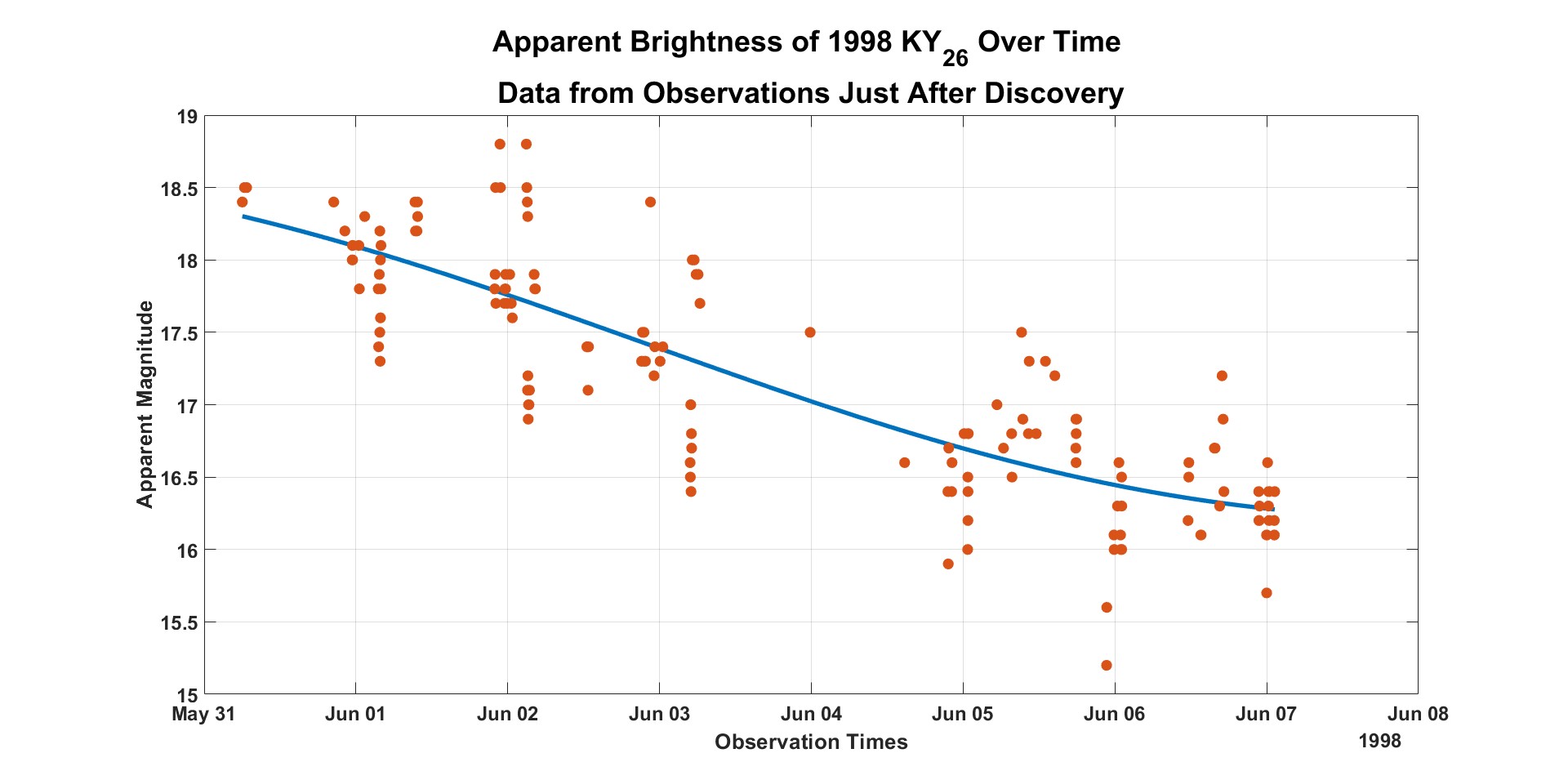}
\caption{Light Curve of 1998 KY$_{26}$ just after discovery. }
\label{fig:Mag}
\end{figure}

The light curve for the short period after discovery in 1998, is given in Figure \ref{fig:Mag}. Observe there are fluctuations in apparent magnitude of $\sim{2.0}$. This is a huge amplitude and very suggestive of a rapidly spinning elongated object.\\

For a projected aspect ratio $a/b$, this is related to fluctuations in apparent magnitude, $\Delta$M, by:
\begin{equation}
    \Delta M= 2.5~log_{10}\left(\frac{a}{b}\right)
\end{equation}
This can be rearranged as follows:
\begin{equation}
  \left(\frac{a}{b}\right) = 10^{\frac{\Delta M}{2.5}}
\end{equation}
This leads to an aspect ratio of approximately 6.3:1, representing a very elongated object such as possibly the Phobos 1 probe (which had extended solar panels).

\subsection{Implications}
The combination of high albedo from \cite{santana2025hayabusa2} of 0.52 $\pm$  0.08, the diameter of 11 $\pm$ 2 \si{m}, and low spin period of 5.3516 $\pm$ 0.0001 \si{mins}, all suggest a small, tough, monolithic object, such as the Phobos 1 probe.

\section{Possible Scenario}

The details of the Phobos 1 mission, as generated by OITS are provided in Table \ref{Velocities}, and a comparison of the orbital parameters between Phobos 1 and 1998 KY$_{26}$, is given in Table \ref{Orbits}.\\

As a historical perspective, following a request from the Moscow 'Space Research Centre', the 'European Southern Observatory' (ESO) attempted to observe the Phobos 1 probe 20 days after mission failure \citep{ESO}. The result was negative in that the probe was not visible at the predicted coordinates, however referring to the article, they suggest this was due to 'a faulty rocket fire after contact was lost?'.\\

Let us pursue this hypothesis and model the suspected thrust as an impulsive change in velocity ($\Delta$V) some time between loss of mission and the non-observation of the probe by ESO on 22/09/1988, a period encompassing 20 days. This impulsive $\Delta$V would be insufficient to align the orbits of the two bodies since, as is apparent from Figure \ref{fig:Orbdepict}, there was a large discrepancy in heliocentric longitude (or alternatively mean anomaly) between the two objects at launch, and therefore at the time of mission loss. Consequently the next simplest scenario can be applied with the assumption that there is a further, second $\Delta$V, some time before 1998 KY$_{26}$ was discovered.\\

To this end some REBOUND \citep{2012A&A...537A.128R} N-body integration software was developed. First, the state vector of the probe while it was en route to Mars (specifically on 02/09/1988, the day of loss of mission), was supplied by examination of OITS trajectory results. The REBOUND simulation was intialized at this point. The state vectors of the Sun, the Moon and all the planets were then extracted at this time by executing the NASA SPICE \citep{ACTON199665,ACTON20189} command \textit{spkezr\_c} with the binary SPICE kernel file \textit{de430.bsp} providing the necessary data. The command line argument to this software included the path of a file containing 8 numbers, as follows:
\begin{enumerate}
    \item The first $\Delta$V component x in 'ECLIPJ2000', $\Delta$V1$_x$
    \item The first $\Delta$V component y in 'ECLIPJ2000', $\Delta$V1$_y$
    \item The first $\Delta$V component z in 'ECLIPJ2000', $\Delta$V1$_z$
    \item The time in days to first $\Delta$V1
    \item The second $\Delta$V component x in 'ECLIPJ2000', $\Delta$V2$_x$
    \item The second $\Delta$V component y in 'ECLIPJ2000', $\Delta$V2$_y$
    \item The second $\Delta$V component z in 'ECLIPJ2000', $\Delta$V2$_z$
    \item The time in days between first and second $\Delta$V    
\end{enumerate}
There were two outputs of this software - first the mean square \textit{Mahalanobis Distance} \citep{mahalanobis1936} between the spacecraft and 1998 KY$_{26}$ as of the epoch 07/04/2000 08:13:52. The reference state vector and covariance matrix (used to derive the Mahalanobis distance) were taken from \textit{find\_orb} \citep{find_orb} at this epoch time. The second output was the sum of the magnitudes of the two $\Delta$Vs with 1.9 \si{km.s^{-1}} subtracted. This was to impose a constraint on the total magnitude of $\Delta$V, to align it with a reasonable level expected from the probe.\\

Two optimization paradigms were utilized, first Global Non-Linear Programming (NLP) software in the form of NOMAD \citep{LeDigabel2011}, followed then by Covariance Matrix Adaptation Evolutionary Strategy (CMA-ES) \citep{hansen2001} as implemented by \cite{pycma}. For an attempt to explain the logic behind CMA-ES intuitively, go to \cite{hansen2023}.\\

The workflow adopted was to first try and hone the independent variables listed above to around a global minimum (NOMAD), following this up by a more structured evolutionary strategy to find the precise minimum using CMA-ES. This combined effort proved very fruitful.\\

In the early stages of this workflow, a scaled displacement in 6D phase space was calculated by the software, and output to the optimizers as an objective to minimize, like so:
\begin{equation}
   D = \sqrt{\left(\frac{DX}{AU}\right)^2+\left(\frac{DV}{AU/day/0.017}\right)^2} 
\end{equation}

where in the above, $DX$ and $DV$ are respectively the distance and velocity difference between the probe and 1998 KY$_{26}$, with the 0.017 figure as a convenient scaling factor. This produced good convergence with a combination of NOMAD followed by CMA-ES.\\

However, with the Mahalanobis distance, $MD2$, as the objective, this only reduced to an acceptable level when a 9$^{th}$ optimizable parameter, $\Delta$T, was added to the 8 enumerated above. The optimal magnitude of this $\Delta$T was minimal at only $\sim{2}$ hours, and might be attributed to a slight error in phase between the two objects, possibly originating from an error in the time-stamp generated by OITS, or other unknown causes.\\

The final solution is provided in Table \ref{SOL}. It is interesting to note that, as already indicated, the total $\Delta$V was constrained to 1.9 \si{km.s^{-1}}, and furthermore no additional optimal solutions with tighter $\Delta$V constraints were forthcoming. It should also be affirmed that, by the nature of global optimization, this is NOT a non-existence proof.\\

It is now necessary to compare this level of $\Delta$V with the total available to the Phobos 1 probe, since the mission was lost very early on, and nearly all allocated propellant would have been present. From \cite{Phobos1_2}, the propulsion system exploited for Mars Orbital Insertion (MOI) was a combination of nitric acid and an amine-based propellant. Such a propulsion system has huge momentum change capability. Let us adopt a specific impulse of I$_{sp}$ $\sim{315}$ \si{secs}, and exploit the famous Tsiolkovsky rocket equation:
\begin{equation}
    \frac{M_f}{M_0}=exp\left(-\frac{\Delta V}{gI_{sp}}\right)
\end{equation}
This leads to a propellant mass fraction of 1-$M_f/M_0$ $\sim{46} \%$.\\

From \cite{sagdeev1990brief}, the autonomous thruster for MOI was 3600 kg compared to the overall mass of the probe of 6200 kg. This represents a mass fraction of at most $\sim{58} \%$, disregarding the structural mass of the autonomous thruster. In conclusion, $\Delta$V = 1.9 \si{km.s^{-1}} would seem to be within the propulsion envelope of the Phobos 1 probe, adding plausibility to the association with 1998 KY$_{26}$.\\

\begin{figure}[hbt!]
\centering
\includegraphics[width=0.9\textwidth]{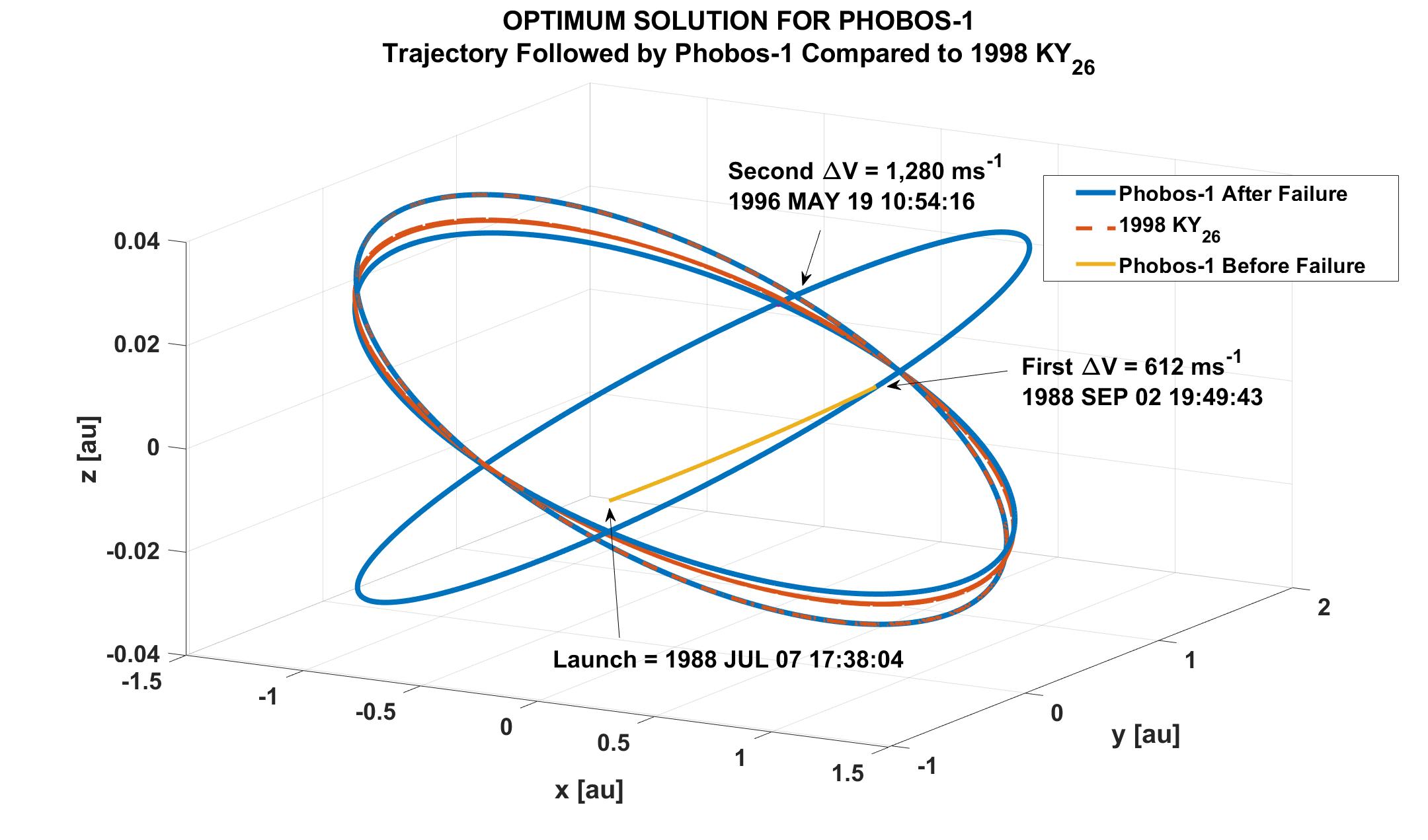}
\caption{Solution trajectory followed by the Phobos 1 probe (blue line) compared to that of 1998 KY$_{26}$ retrodicted by NASA Horizons, SPICE (red dashed line)}
\label{fig:RefOrb}
\end{figure}

\begin{figure}[hbt!]
\centering
\includegraphics[width=0.9\textwidth]{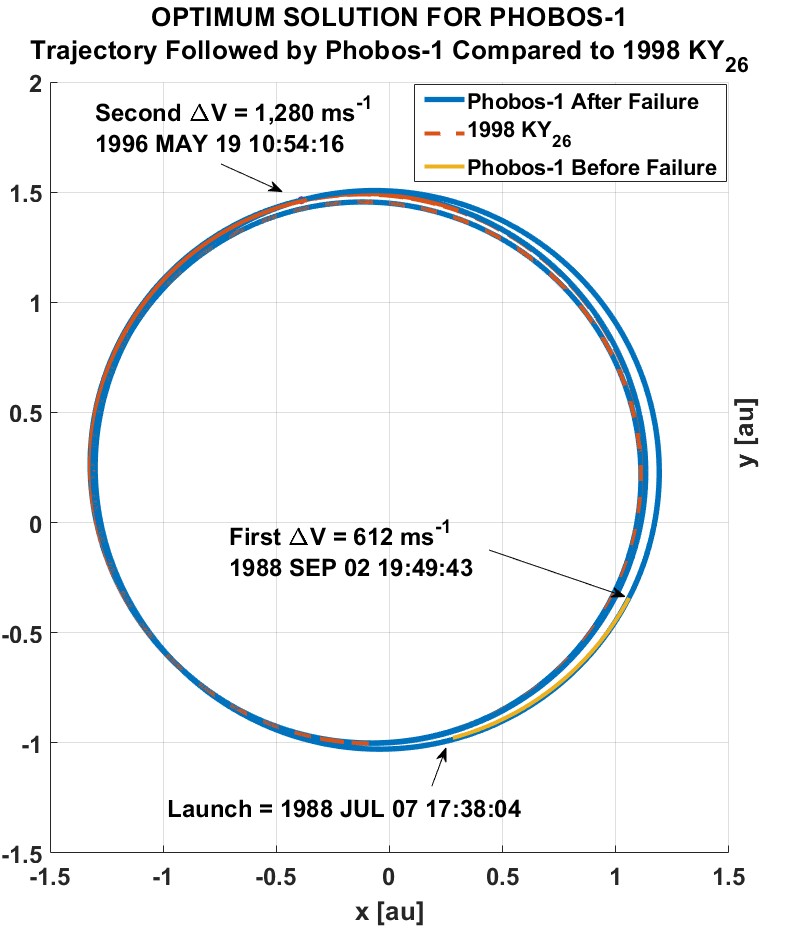}
\caption{Solution trajectory (plan view) followed by the Phobos 1 probe (blue line) compared to that of 1998 KY$_{26}$ retrodicted by NASA Horizons, SPICE (red dashed line)}
\label{fig:RefPlan}
\end{figure}

\begin{figure}[hbt!]
\centering
\includegraphics[width=0.9\textwidth]{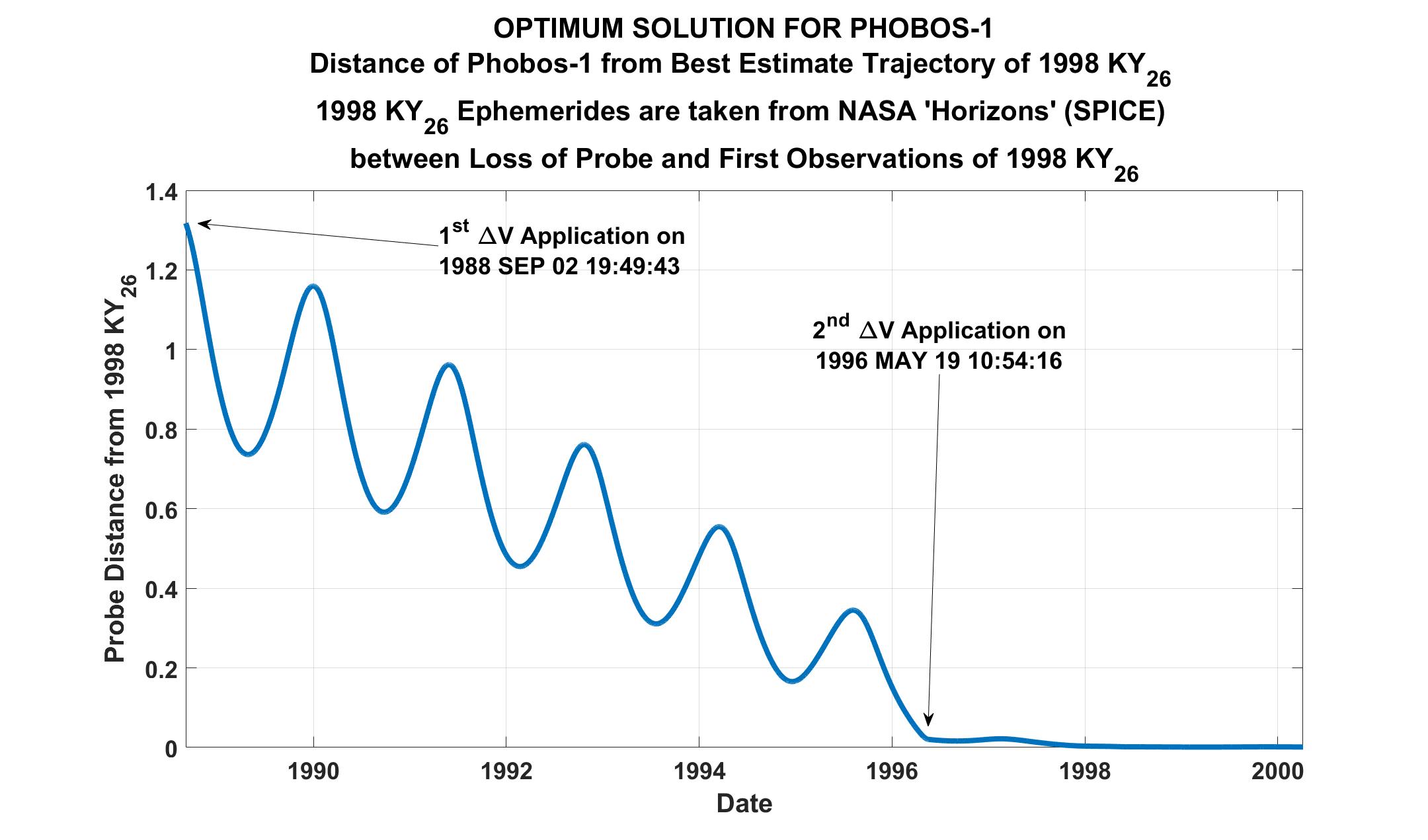}
\caption{Displacement between the Phobos 1 probe and 1998 KY$_{26}$}
\label{fig:DISP}
\end{figure}

Attention is now drawn to the orbital path followed by the Phobos 1 probe according to the solution summarised in Table \ref{SOL} - refer Figure \ref{fig:RefOrb} for the 3D perspective and the plan view is provided in Figure \ref{fig:RefPlan}. The displacement between the positions of the two bodies is shown in Figure \ref{fig:DISP}.\\

An animation by Tony Dunn of this orbital path is available \citep{orb_sim}.

\begin{table}[]
\centering
\begin{tabular}{|c|c|c|ccc|}
\hline
\textbf{}          & \textbf{Optimized} & \textbf{}      & \textbf{}       & \textbf{}           & \textbf{}           \\
\textbf{Parameter} & \textbf{Parameter} & \textbf{Units} & \textbf{Notes:} & \textbf{}           & \textbf{}           \\ \hline
$\Delta$V1$_x$ & -0.579 & \si{km.s^{-1}}  &      &          &      \\ \cline{1-3}
$\Delta$V1$_y$ & -0.049 & \si{km.s^{-1}}   & mag: & 0.612096 & km/s \\ \cline{1-3}
$\Delta$V1$_z$ & 0.191  & \si{km.s^{-1}}   &      &          &      \\ \hline
Time $\Delta$V1    & 0.826              & days           & Date:           & \multicolumn{2}{c|}{1988 SEP 02 19:49:43} \\ \hline
$\Delta$V2$_x$ & 0.010  & \si{km.s^{-1}}   &      &          &      \\ \cline{1-3}
$\Delta$V2$_y$ & 0.835  & \si{km.s^{-1}}   & mag: & 1.279599 & km/s \\ \cline{1-3}
$\Delta$V2$_z$ & 0.970  & \si{km.s^{-1}}  &      &          &      \\ \hline
Time $\Delta$V2    & 2815.628           & days           & Date:           & \multicolumn{2}{c|}{1996 MAY 19 10:54:16} \\ \hline
$\Delta$T      & 2.010  & hours &      &          &      \\ \hline
\end{tabular}
\caption{Optimal Solution found by NOMAD and then CMA-ES, ultimately attaining an insignifcantly small square Mahalanobis distance}
\label{SOL}
\end{table}

\section{Discussion}
To be clear, this research cannot unequivocally identify that 1998 KY$_{26}$ is really the Phobos 1 probe, it analyses the evidence in support of this conclusion, which is quite compelling. We have found:
\begin{enumerate}
    \item The Phobos 1 and 1998 KY$_{26}$ orbits are visually similar
    \item The difference between these orbits is compatible energetically with the overall $\Delta$V envelope available to Phobos 1
    \item The 2 orbits converge and are statistically compatible, given the uncertainty in the orbit of 1998 KY$_{26}$, which is tightly constrained due to the existence of over 230 observations of this dark comet
    \item There is a historical record in support of the hypothesis that a propulsive $\Delta$V was delivered shortly after loss of mission
    \item The Phobos 1 mission was lost early on in the probe's transit to Mars enabling a large $\Delta$V capability
    \item There is supportive observational data of the dark comet, i.e. size, albedo and spin
    \item The object appears to be quite elongated from changes in its apparent magnitude
\end{enumerate}

This can be compared to the evidence to the contrary:
\begin{enumerate}
    \item it is generally assumed by the scientific community that this is a natural body
    \item there are AI-assisted (SAGE) reconstructions of the object based on optical and RADAR data, which suggest it is an asteroid with a random asteroid-type shape, though with apparent 'concavities' \citep{SantanaRos2025}
\end{enumerate}

So to the question: 'is 1998 KY$_{26}$ the Phobos 1 probe?' there is plenty of supporting evidence, yet the JAXA Hayabusa2 mission to this object in 2031 will effectively 'put the cat amongst the pigeons'.  

\section{Conclusion}
This paper identified several independent lines of evidence that motivate further consideration of the possibility that the dark comet 1998 KY$_{26}$ is in fact technogenic, specifically the Russian Phobos 1 probe to Mars. It suggests a particular historical path through which the trajectories of the probe and the dark comet could evolve to become effectively statistically indistinguishable. In anticipation of the Hayabusa2 observations in 2031, which will be decisive in resolving the origin of this object, we encourage further observational, dynamical, and theoretical studies aimed at more tightly constraining the nature and properties of 1998 KY$_{26}$.

\section{Acknowledgments}

AI (ChatGPT) was used to assist the research. Thanks to John I. Davies (i4is) who provided various important insights also. The orbit simulation \citep{orb_sim} was created by Tony Dunn.

\bibliography{1998KY26}{}
\bibliographystyle{aasjournalv7}



\end{document}